\documentclass[aps,prb,twocolumn,showpacs,amsmath,amssymb,superscriptaddress, citeautoscript]{revtex4-2}
\setcitestyle{super}
\usepackage{graphicx}
\usepackage{hyperref}
\usepackage{xcolor}
\usepackage{color}
\usepackage[all,cmtip]{xy}
\usepackage{multirow}
\usepackage{natbib}
\usepackage{amsmath,amsthm} 
\usepackage{amsfonts}
\usepackage{soul}

\newcommand{\minus}{\scalebox{0.5}[1.0]{$-$}}
\makeatother

\begin{document}

\title{Spin Weyl Topological Insulators}

\author{Rafael Gonz\'{a}lez-Hern\'{a}ndez}
\email{rhernandezj@uninorte.edu.co}
\affiliation{Departamento de F\'{i}sica y Geociencias, Universidad del Norte, Km. 5 V\'{i}a Antigua Puerto Colombia, Barranquilla 080020, Colombia}
\author{Bernardo Uribe}
\email{bjongbloed@uninorte.edu.co}
\affiliation{Departamento de Matem\'{a}ticas y Estad\'{i}stica, Universidad del Norte, Km. 5 V\'{i}a Antigua Puerto Colombia, Barranquilla 080020, Colombia}

\date{\today}

\begin{abstract}
	The quantum nature of electron spin is crucial for establishing topological invariants in real materials.
	Since the spin does not in general commute with the Hamiltonian, some of the topological features of the material can be extracted from
its study. In insulating materials, the spin operator induces a projected
	operator on valence states called the spin valence operator. Its spectrum 
	contains information with regard to the different phases of the spin Chern class.
	If the spin valence spectrum is gapped, the spin Chern numbers are constant along parallel planes thus defining spin Chern insulating materials. If the spin valence spectrum is not gapped, the changes in the spin Chern numbers occur whenever this spectrum is zero.
	Materials whose spin valence spectrum are gapless will be denoted spin Weyl topological insulators and its definition together with some of their properties will be presented in this work. The classification of 
materials from the properties of the spin valence operator  provides
	a new characterization which complements the existing list of topological invariants.

\end{abstract}
\maketitle

\section{Introduction}

Topological insulators exhibit a unique electronic structure where the bulk of the material remains insulating due to the presence of a large energy band gap, while the surface or edge states emerge when the material interfaces with a trivial insulator \cite{Colloquium-topological-insulators,Topological-insulators-and-superconductors}.
This feature of a material can be related to classical topological invariants associated to the vector bundle of occupied states (valence eigenstates)
such as the first Chern number in the 2D case and the Chern-Simons invariant ($\theta$-term) in the 3D case \cite{TKNN-invariant,Xiao-Hughes-Zhang, Magnetoelectric-Vanderbilt}.

The characterization of the topological nature of a material from first-principles calculations has been extensively studied, and several 
procedures have been established to extract its features, such as Wilson loop calculations \cite{Wilson-loop-ti,Wilson-loop-fragile}, eigenvalues of crystal symmetry operators \cite{irrep,Tang-symm-indicators} and elementary band representation of valance bands \cite{Topological-quantum-chemistry,Magnetictopologicalquantumchemistry}, among others.  
In this work we propose an alternative strategy to detect the topological nature of a material which is based on the topological properties of the spin operator. The 
noncommutativity of the spin operator with the Hamiltonian permits to infer relevant information
of the material 
from the spectrum of the spin operator once it is restricted to the valence bands \cite{Prodan-SCN}.
The geometrical and topological properties of this spectrum follow the same structural behavior
as the ones from the energy spectrum  \cite{Topological-semimetals,Phases-Weyl-semimetals}. 
Gapless systems will then have Weyl points (spin
Weyl points),  and the understanding of the location and the chirality of these points is the key ingredient 
underlying our proposal for a new indicator.

Our proposal for an indicator enhancing the known classification of topological materials is called the spin invariants vector. This vector has seven integer numbers where the first 
number counts the number of spin Weyl points with positive chirality and the next six
are the Chern numbers of the negative spin valence eigenvalues across the planes $k_l$=0, $\pi$ for $l$=$x,y,z$. The first number is denoted spin Weyl indicator and it is zero only when the spin valence
spectrum is gapped. In this case the spin Chern numbers are constant across parallel planes and the 
next six coordinates simply encode these constant Chern numbers; in this case the material is a spin Chern insulator when some Chern number is non zero, or a spin insulator if all Chern numbers are trivial.
If the spin Weyl indicator is not zero, then the spin valence spectrum is gapless and the Chern number
across parallel planes may differ, we call these materials spin Weyl topological insulators. The next six coordinates provide the information of the Chern numbers across the six planes $k_l$=0, $\pi$, and the spin Weyl indicator measures the total positive change
of Chern number across parallel planes.

The parity of the spin Weyl indicator is precisely the Fu-Kane-Mele invariant, and in the case 
of the spin Weyl indicator being even, it provides an enhancement for the detection of weak and fragile topological phases \cite{Fragile-Topology}.

In order to understand the features of the spin invariant vector, we analize 
the behavior of the spin valence operator in the 3D Bernevig-Hughes-Zhang (BHZ) model \cite{BHZ}, 
as well as in particular tight-binding Hamiltonian \cite{Vanderbilt-axion}.
In the 3D BHZ model we obtain five different topological phases depending on one parameter. 
Two trivial insulator phases, two strong topological insulator phases and one spin Weyl topological
insulator phase. This last phase exhibits a gapless spin valence spectrum with a total of four spin Weyl points. Its topological nature is also inferred from the change of spin Chern number from -1
in $k_z=0$ to 1 in $k_z=\pi$ (see Fig \ref{phases BHZ}). The tight-binding Hamiltonian exhibits
a strong topological insulator phase with two Weyl points  and with spin invariant vector $(1|10\minus1000)$. Finally, we calculate the properties of the spin valence spectrum in materials Bi$_2$Te$_3$, Bi and SnTe, thus
characterizing Bi$_2$Te$_3$ as a strong topological insulator, Bi as a spin Chern insulator
and SnTe as a spin Weyl topological insulator. All three materials exhibit a non-zero spin Hall Effect, 
suggesting that the appearance of the phenomenon is predicted by  the non-triviality of the spin Chern numbers.

The spin Chern numbers thus become a customized tool for the distinction and classification of topological insulators. This ansatz was originally put forward by Prodan\cite{Prodan-SCN}
who showed the robustness of the spin Chern numbers and carried out an extensive analysis of the properties
of the spin valence spectrum. The implications of the robustness of the spin Chern numbers have been explored by other
authors \cite{spin-texture-fragile}, and in the case of 3D insulating materials, a comprehensive analysis of both theoretical and computational aspects has been studied by several authors \cite{chern-number-3d-ti,spin-resolved-3d}. 
By introducing nested spin-resolved Wilson loops and layer constructions,   
Lin et al. \cite{spin-resolved-3d} studied the behavior of the spin Chern numbers across parallel planes in the BZ, thus describing topological properties of these numbers while detecting
the presence of spin Weyl points. This analysis allowed Lin et al. \cite{spin-resolved-3d} to propose
a novel method for further classifying topological insulating phases.

In the present paper we further explore the properties of the spin valence spectrum in 3D insulating materials and we propose
a new indicator that can effectively identify 3D topological insulating phases.
This newly established indicator, based on the spin properties, offers valuable insights into the fundamental understanding and potential applications of topological materials.

\section{Invariants of Topological Insulators}

The mathematics behind the theory of topological invariants in insulating materials has been extensively
studied \cite{vanderbilt_book,moore_book}. Several methods for detecting topological invariants of a prescribed Hamiltonian
have been established and the construction of a comprehensive list of all possible indicators for such invariants is currently an active area of research \cite{Catalogueoftopologicalelectronicmaterials,Acompletecatalogueofhigh-qualitytopologicalmaterials}. 

The importance of these topological invariants lies in the amazing relation that some of them have with certain electromagnetic properties of materials. For instance, in 2D insulators, the Chern number of the occupied states 
provides the quantization of the Anomalous Hall Effect \cite{Quantized-Hall-conductance}, and in 3D insulators with Time Reversal Symmetry (TRS), the non-triviality of the Fu-Kane-Mele invariant (FKM) provides an indicator of a strong topological insulating type \cite{Fu-Kane-Mele}; just to mention a few.

The main line of thought underlying the existence of the topological invariants in insulators goes as follows.
The Hamiltonian of the periodic system provides a Hermitian operator acting on parametrized vectors over the Brillouin Zone (BZ):
\begin{align}
\widehat{H} : \Gamma(\mathbb{C}^N \times B) \to \Gamma(\mathbb{C}^N \times B),
\end{align} 
here $N$ is the number of bands, $B$ denotes the BZ and $\Gamma$ denotes
the space of sections of the trivial complex vector bundle $\mathbb{C}^N \times B$.

Whenever there is an energy gap at the Fermi level on the eigenvalues of the Hamiltonian, we
say that the material is insulating. The insulating condition permits to separate the  valence states  from the conducting ones. This separation at the level of vector bundles
defines the partition
\begin{align}
\mathbb{C}^N \times B \cong E^{val} \oplus E^{cond}
\end{align}
where the sections of the bundle $E^{val}$ are generated by the valence
eigenvectors of the Hamiltonian $\{ | \psi_i \rangle \}_{i=1}^{n_{occ}}$,
where $n_{occ}$ indicates the number of occupied bands which is the rank of $E^{val}$.

Since $E^{val}$ is a complex vector bundle over the BZ, we may assign to it
the topological invariants that it defines in the complex K-theory groups. 
The only interesting invariant  that appears here, besides the rank of the vector bundle, is the first Chern class.
This first Chern class $c_1:=c_1(E^{val})$ can be evaluated on planes inside the BZ and the 
associated numbers can be determined.

If the first Chern class $c_1$ is not zero, the material is called a Chern insulator.
In the 2D case it provides the quantization of the Anomalous Hall Effect and several
materials exhibit this property \cite{tm_qahe}. On the other hand, 3D materials
with a Chern Insulating property have been elusive to detect and until now none single 3D material exhibits this property. 

By incorporating geometrical and physical symmetries of the Hamiltonian into the analysis, more
specific information regarding the topological invariants can be deduced. If the system preserves TRS ($\mathbb{T}$), and we are in the spin orbit coupling (SOC) environment with $\mathbb{T}^2=-1$, then 
the FKM invariant provides an indicator of the strong topological insulating property \cite{Fu-Kane, Fu-Kane-Mele}. If the symmetry preserved is $C_2 \mathbb{T}$, a combination of a $180^o$ rotation and TRS,
then an indicator of being an axion insulator is the Stiefel-Whitney class of the $C_2 \mathbb{T}$ invariant real
vector bundle of $E^{val}$ restricted to the planes fixed by $C_2 \mathbb{T}$ \cite{Axion_insulators_GPU}.

It is important to notice that every extra crystal symmetry will induce topological invariants.
Sometimes the invariants already appeared due to another symmetry, but some other times the
invariant is new and may or may not imply the existence of invariants of other symmetries.
The task of finding a complete set of indicators for all geometrical symmetries is an ongoing subject of 
research.

Besides the geometrical symmetries, there are also the physical symmetries. These are the ones
that come from the fact that we are dealing with a quantum mechanical system.
One such symmetry is the spin, and incorporating it into the analysis of topological invariants
has been very fruitful\cite{Prodan-SCN}. In what follows we will study some of the topological invariants which can be extracted when the spin operator acts in the occupied wave function space.

\section{Spin Weyl Indicator}

If the spin operator $\widehat{S}_z$ does not commute with the Hamiltonian, we cannot expect to simultaneously diagonalize it with the Hamiltonian.
In order to obtain a symmetry of the vector bundle $E^{val}$, we compose the action of the spin operator
with the projection onto the occupied states:
\begin{align}
\widehat{S}_z^{val} : \Gamma(E^{val}) &\to \Gamma(E^{val})  \\
\varphi  &  \mapsto \pi^{val} \circ \widehat{S}_z (\varphi);
\end{align}
here $\pi^{val} : \Gamma(\mathbb{C}^N \times B ) \to \Gamma(E^{val})$ is the projection
operator $\sum_{i=1}^{n_{occ}}| \psi_i \rangle \langle \psi_i |$.

We call the operator $\widehat{S}_z^{val} $ the spin valence (SV) operator and one could see it as
a physical symmetry of the bundle $E^{val}$ of occupied states. 
This SV operator can be diagonalized, and its spectrum can be studied with exactly the same
tools as the ones used to understand the spectrum of the Hamiltonian.
We have used the $\widehat{S}_z^{val}$ component for the spin operator, but it can be generalized to include other components.

Note that the SV spectrum takes values in the interval $[-1,1]$ (in $\hbar$/2 units), and moreover, that an eigenvalue of zero means that there is a combination of occupied states whose spin lies completely 
on the unoccupied states. Hence it is important to measure whether the SV spectrum crosses the
zero value or not, and two different scenarios appear (See Figure \ref{classes}).

\subsection{Gapped Spin valence spectrum}

If the SV spectrum is gapped at zero, then we can partition the vector bundle $E^{val}$ into
two different vector bundles: 
\begin{align}
E^{val} \cong E^{val}_{s_z^+} \oplus E^{val}_{s_z^-}
\end{align}
where $E^{val}_+$ and $E^{val}_-$ denote respectively the positive and negative 
SV eigenstates.

The topological invariants associated to these two complex vector bundles is the first Chern classes
\begin{align}
c_1^{s_z^\pm}:=c_1(E^{val}_{s_z^\pm})
\end{align}
and since 
\begin{align}
c_1=c_1^{s_z^+}+c_1^{s_z^-}, 
\end{align}
 the new invariant is usually defined
as half the difference between the two Chern classes:
\begin{align}
c_1^{s_z}= \tfrac{1}{2} \left(c_1^{s_z^+}-c_1^{s_z^-}\right).
\end{align}
This Chern class is called the spin Chern class, and together with the total first Chern class
$c_1$, determines uniquely the first Chern class of both the positive and the negative 
SV states.

When TRS is preserved, the vector bundles  $E^{val}_{s_z^+}$ and $E^{val}_{s_z^-}$ are isomorphic
via the antiunitary transformation defined by $\mathbb{T}$. In this case $c_1^{s_z^+}=-c_1^{s_z^-}$
and therefore the spin Chern class $c_1^{s_z}$ is the invariant preserved. In 2D materials 
whose spin almost commutes with the Hamiltonian, the spin Chern number
is a good indicator for the Quantum spin Hall Effect. These materials
are called quantum spin Hall insulators (QSHI) and they include both functionalized and pristine antimonene and bismuthene 2D materials \cite{Bay-antimonene,Wang-antimonene}.

In 2D systems Chern classes are uniquely determined by the Chern number, namely
the integration of the Chern class on the whole BZ. While on 3D systems they are determined
by the integrals along all closed surfaces in the BZ. The value of this integral over any closed surface $surf$ is the Chern number associated with the surface:
\begin{align}
c_1^{s_z^\pm}(surf) := \int_{surf} c_1(E^{val}_{s_z^\pm}).
\end{align}

The spin Chern numbers are usually associated with planes of the form $k_l$=0,$\pi$,
and in the case of a gapped SV spectrum, these numbers are constant along parallel planes. 
The spin Chern number can only vary on parallel planes whenever the SV spectrum is gapless,
thereby indicating the presence of spin Weyl (SW) points.

\subsection{Gapless Spin valence spectrum}

Whenever the SV spectrum is not gapped, we cannot partition the occupied states
into positive and negative SV eigenvectors all across the BZ. But outside the points
in momentum space where the SV eigenvalue is zero, this partition can be performed.

Call  {\bf  Spin Weyl (SW) points}  the points in momentum space where there is a  zero SV eigenvalue.
Around each SW point $\bf{k}$, a 2D sphere $S_\epsilon({\bf{k}})$ of small radius $\epsilon >0$ could be defined. The vector
bundle of occupied states restricted to this sphere splits into positive and negative SV
eigenvectors. Therefore we could associate to the SW point the Chern number of the negative SV
eigenvectors restricted to the sphere. 

Mimicking the definition of the chirality of Weyl points 
of the Hamiltonian, we define 
 the {\bf spin Chirality} of the SW point $\bf{k}$ as follows:
\begin{align}
\chi^{s_z}({\bf{k}}) := c_1^{s_z^-}(S_\epsilon({\bf{k}})).
\end{align}

By the Nielsen-Ninomiya Theorem \cite{Absence-of-neutrinos-in-lattice}, the sum
of the spin Chiralities of all SW points is zero:
\begin{align}
\sum_{{\bf k} \in SW} \chi^{s_z}({\bf{k}})  =0,
\end{align}
where $SW$ denotes the finite set of spin Weyl points in a generic Hamiltonian.

Therefore the maximum Berry curvature flux of the negative SV states 
is given by the sum of all positive spin chiralities.  Simply enclose with a closed surface
the spin Weyl points with positive chirality, and by Gauss' law, the total flux is the sum of
the chiralities inside the surface.

We claim that this maximum Berry curvature flow of the negative SV states 
is a topological indicator of the system. We therefore propose to define the
{\bf spin Weyl indicator} (SWI) of the system as the sum of the positive chiralities of all SW points:
\begin{align}
\mathrm{SWI}: = \frac{1}{2}\sum_{{\bf k} \in SW} |\chi^{s_z}({\bf{k}})|.
\end{align}
The formula above simply computes the sum of the absolute value of all chiralities and divides by two.
This way there is no need to distinguish the positive chiralities from the negative ones.

The SWI is a natural number that encodes topological information about the system.
For instance, in the presence of TRS or TRS composed with a 2-fold rotation, the parity of the SWI is respectively equivalent to the FKM invariant \cite{Fu-Kane, Fu-Kane-Mele} or to the value of the $\theta$-term\cite{Axion_insulators_GPU}.
Whenever the SWI is zero, the projected spin operator is gapped, and the material
can be classified as spin Chern insulator if any spin Chern number is not zero. 
Whenever SWI is even, the material might be endowed with weak or fragile topological phases.

Calculating the SWI by detecting the SW points together with their
chiralities might be cumbersome. Alternatively, we propose to calculate the Chern number
of the negative SV eigenstates 
across planes perpendicular to a given axis. By choosing the $k_l$-axis for $l=x,y,z$ we  
calculate and plot the function
\begin{align}
\mathrm{SCN}: [-\pi, \pi] \to &\mathbb{Z} \\
t \mapsto &
c_1^{s_z ^-}( k_l=t).
\end{align}
This function increases and decreases by integer values whenever SW points are crossed, and the total amount of positive 
changes is precisely the SWI. We will call this function the signal of the  {\bf Spin Chern number (SCN)}; cf. \cite{chern-number-3d-ti}

The SCN signal has been plotted in the 3D Bernevig-Hughes-Zhang (BHZ) model in Fig. \ref{phases BHZ},
in the Tight-Binding  (TB) model of pristine pyrochlore in Fig. \ref{modelo-tb}c) and in
 real materials Bi$_2$Te$_3$ and SnTe in Fig. \ref{modelo-bi2se3} c) and c') respectively.
 In all these cases the value of the SWI can be deduced from the signal of the SCN in the BZ.
 
 \subsection{Spin invariants vector}
 
 It is important to notice that the SWI cannot be deduced solely from the 
 spin Chern numbers of the planes $k_l=0,\pi$ with $l=x,y,z$. The material
 SnTe has spin Chern number equal to 2 along the planes $k_l=0, \pi$ for $l=x,y$
 and 0 along the planes $k_z=0,\pi$ (see Fig. \ref{modelo-bi2se3} c')), while
 its SWI equals 8. 
 
 A coherent set of invariants associated with the SV operator should therefore include
 the spin Chen numbers along the preferred planes plus the value of the SWI. Hence
 we propose to define the {\bf spin invariants vector} as the array of seven integer numbers
 \begin{align}
 (n|n_x^0,n_x^\pi,n_y^0,n_y^\pi,n_z^0,n_z^\pi)
 \end{align}
where $n=\mathrm{SWI}$ and $n_l^w=c_1^{s_z^-}(k_l=w)$. 

\begin{figure}
	\includegraphics[width=8.2cm]{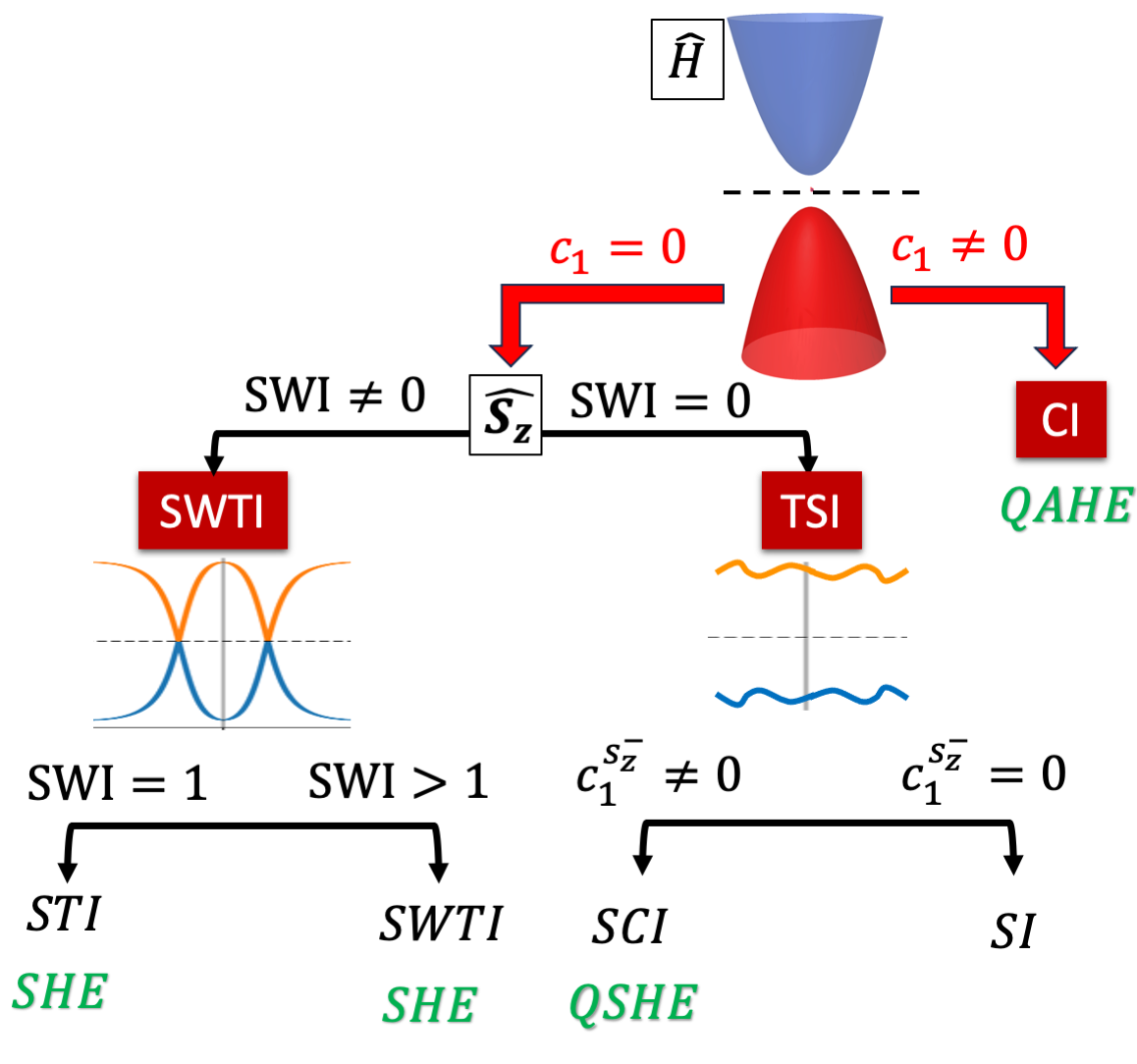}
	\caption{Topological classification of 3D insulators using Chern classes and the SWI (spin Weyl indicator).If the total Chern class $c_1$ of the valence states is non-trivial, there
	is a plane where the Chern number is not zero. In this case the material is a Chern insulator (CI) and the anomalous Hall effect is quantized (QAHE). When the total Chern class of the valence states is trivial, then the SWI number indicates whether the spin Chern numbers vary along parallel planes. Whenever the SWI is trivial, then there are no SW points and therefore the spin Chern numbers do not vary; these materials will be called topological spin insulators (TSI). In this case if the spin Chern class $c_1^{s^{-}_z}$ is trivial, the material is a spin insulator (SI), otherwise the material is a spin Chern insulator (SCI)
and has a quantized spin Hall effect (QSHE).	. 
Whenever the SWI is not zero, the spin Chern numbers vary along parallel planes and therefore the spin Chern numbers along the planes $k_l=0,\pi$ do not constitute a complete topological indicator of the material. The case of the SWI being 1 correspond with most models whose FKM invariant is non-trivial; these are the strong
topological insulators (STI)s. Materials whose SWI is non-trivial will be denoted spin Weyl topological insulators (SWTI); these materials show a spin Hall effect (SHE) inside the energy band gap.}
	\label{classes}
\end{figure}

The spin invariants vector will provide the information necessary to determine the
spin topological classification presented in Fig. \ref{classes}. The zero vector
represents
 a spin insulator, a vector with $n=0$, which is moreover non-trivial, represents a spin Chern insulator, and a vector with $n\neq 0$  represents a spin Weyl topological insulator.

\section{3D BHZ model}  
The 2D Bernevig-Hughes-Zhang (BHZ) two band model (upper left $2 \times 2$
matrix of \eqref{BHZ Hamiltonian}) for a 
spin topological insulator \cite{BHZ} can be used to construct
a four band Hamiltonian in 3D where two opposite copies of the 
2D BHZ model are superposed (matrix of \eqref{BHZ Hamiltonian} with $D=0$).
 In order to obtain
a change of phase for the 2D layers, an off-diagonal term $D$ depending on $k_z$
is added thus obtaining one version of the  3D BHZ
Hamiltonian 
\begin{equation} \label{BHZ Hamiltonian}
H_{BHZ}({\bf k}) =
\left(\begin{matrix}
M & A & 0 & D \\
A^* & -M & D & 0 \\
0 & D &  M  & - A^*\\
D & 0 & -A &-M 
\end{matrix}\right)
\end{equation}
\noindent where
\begin{align}
M=& M_0 -B_0\left(\cos(k_x) + \cos(k_y)+\cos(k_z)\right),\\
A = & A_0 \left( \sin(k_x) + i \sin (k_y) \right),\\
D = & D_0 \sin(k_z).
\end{align}

This Hamiltonian is written in a basis given by the states $|F \!  \uparrow  \rangle, |H \!  \uparrow \rangle, |F \! \downarrow \rangle, |H \! \downarrow \rangle$, in that order, and it models a phase transition 
of a 2D topological insulator.  In what follows we will show that changes in the value of $\frac{M_0}{B_0}$ induce five different insulating phases for this 3D model, 
where each phase change is marked by the closure of the energy band gap.

\begin{figure*}
	\includegraphics[width=16.2cm]{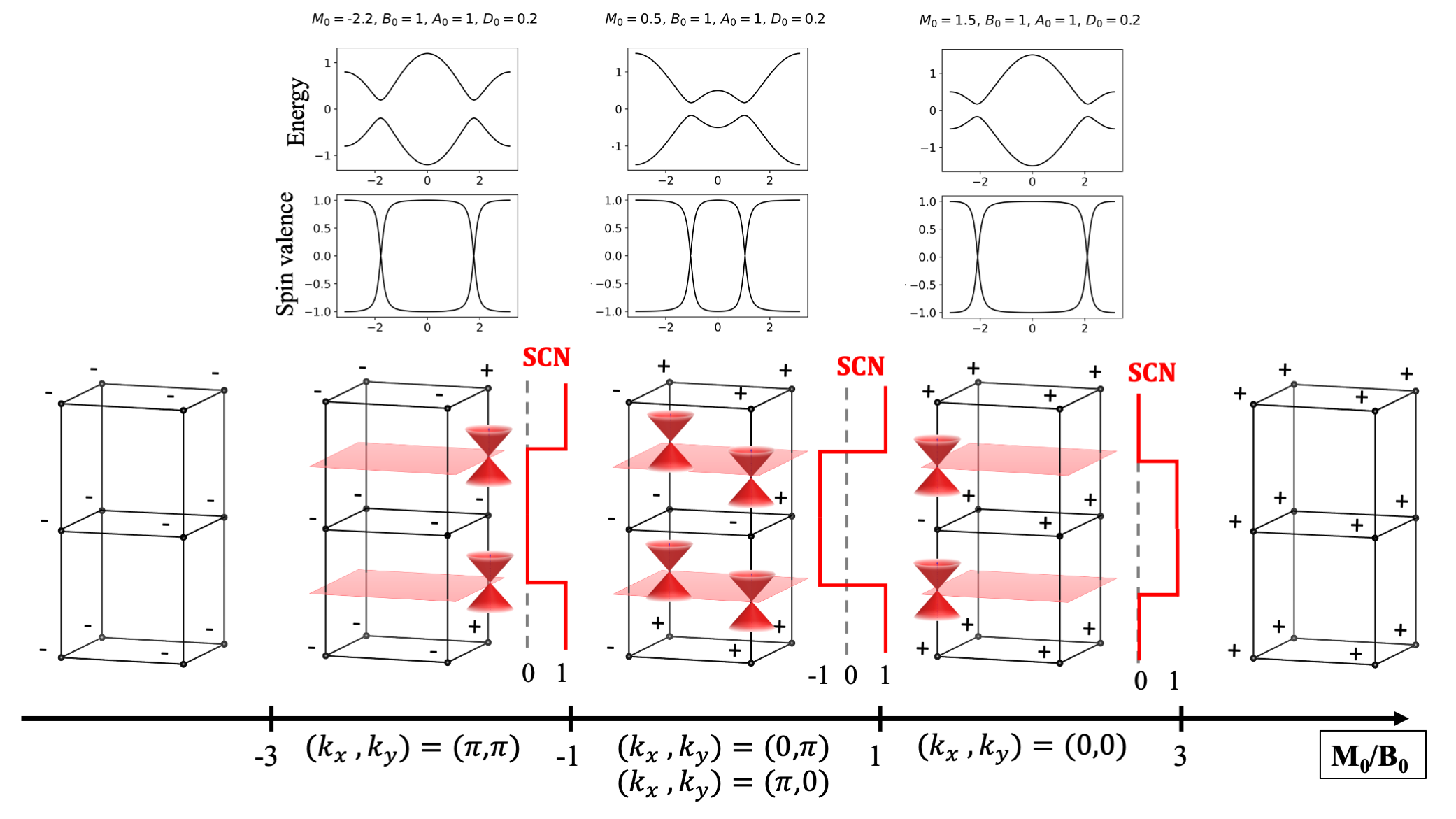}
	\caption{ Five  phases of the 3D BHZ Hamiltonian of Eqn. \eqref{BHZ Hamiltonian}.
For $\tfrac{M_0}{B_0}=-3,-1,1,3$ the Hamiltonian is gapless, and for other values of $\tfrac{M_0}{B_0}$
the Hamiltonian is gapped. For $|\tfrac{M_0}{B_0}| > 3$ the system is a trivial insulator, 
for $1 <|\tfrac{M_0}{B_0}| < 3$ it is a Strong TI with SWI=1 and for $|\tfrac{M_0}{B_0}|  <1$ it is a SWTI with  SWI=2.  
The cones are located where the spin valence eigenvalues are zero, and this
occurs  along axis parallel to the $k_z$ axis: in the first non-trivial phase they are on $(k_x,k_y)=(0,0)$,
on the second non-trivial phase on $(k_x,k_y)=(0,\pi)$ and $(k_x,k_y)=(\pi,0)$, and on the third non-trivial
phase on $(k_x,k_y)=(\pi,\pi)$. The signs on the TRIMs denote the eigenvalues of the inversion operator
on the negative spin valence eigenstate.  The step functions on the right of each BZ is the plot
of the Chern number of the negative spin valence eigenstate on the planes $k_z=t$ when $t$ varies 
from $-\pi$  to $\pi$.  The Energy and spin valence plots vs. the appropriate axis parallel to the
$k_z$-axis in the three non-trivial phases are shown for specific choices of structural constants. The spin Weyl points can be seen in all of them.} 
	\label{phases BHZ}
\end{figure*}

For $3<|\frac{M_0}{B_0}| $ it is a trivial insulator, for $1<|\frac{M_0}{B_0}| <3$ we have an STI whose SWI is 1, and for $-1<\frac{M_0}{B_0} <1$ we have SWTI whose SWI is 2. These phases are illustrated in Figure \ref{phases BHZ}.

The degenerate Eigenvalues of the Hamiltonian are $E=\pm \lambda$ with
\begin{align}
\lambda=
 \sqrt(M^2+|A|^2  + D^2),
\end{align}
and one choice of  eigenvectors are:
\begin{align}
\nu_1 =  &  \left(\left( M - \lambda\right), A^*,0,D \right)  \label{nu1}   \\
\nu_2 =  &  \left(   A,   -\left( M +\lambda\right), D,0 \right) \label{nu2} \\
\nu_3 =  &  \left(\left( M +\lambda\right), A^*,0,D \right) \label{nu3}\\
\nu_4 =  &  \left(  A,   -\left( M - \lambda\right), D,0 \right). \label{nu4}
\end{align}
The energy spectrum is gapless only when $M=A=D=0$ and this only happens
whenever $\frac{M_0}{B_0}= -3,-1,1,3$. For any other choice of $\frac{M_0}{B_0}$
the energy spectrum is gapped.

Note that the valence states $\nu_1$ and $\nu_2$ are not linearly independent as presented above; nevertheless this simple presentation of the eigenvectors permit us to deduce an important result that it will be outlined in what follows.

\subsection{Spin Weyl points}

The spin  operator $\widehat{S}_z$ in this case is the diagonal matrix $\mbox{diag}(1,1,-1,-1)$, 
and we may restrict the spin operator only to the valence states.
If $\psi_j$ are the eigenvectors of the Hamiltonian forming a unitary base (norm one and
perpendicular to one another), then the SV matrix is defined as follows:
\begin{align}
	(M_{{s}_z})_{ij}= \langle \psi_i|\widehat{S}_z(\psi_j) \rangle\ \ i,j \in \{1,2\}.
\end{align}

The eigenvalues of the SV matrix give us the SV spectrum. Whenever
there is an SV eigenvalue gap, we could separate the positive states from the negative
states, and we could find the topological invariants for each group of SV eigenstates.
Where the SV spectrum is not gapped, a spin Chern number transition occurs in the BZ.
Let us show that this indeed is what happens in the 3D BHZ Hamiltonian.

The  SV eigenvalues vanish whenever the whole SV
matrix vanishes. Note that in this case, we could use the degenerate basis $\{\nu_1,\nu_2\}$ 
of Eqns. \eqref{nu1} and \eqref{nu2} in order to solve the equations
\begin{align}
 \langle \nu_i|\widehat{S}_z(\nu_j) \rangle\ = 0 \ \  \ i,j \in \{1,2\}.
\end{align}
The equations become 
\begin{align}
 \langle \nu_1 |\widehat{s}_z(\nu_2) \rangle\   &= 2 A \lambda =0  \label{spinWeylpoint1} \\
 \langle \nu_2 |\widehat{s}_z(\nu_1) \rangle\   &= 2 A^* \lambda =0  \label{spinWeylpoint2}\\
\langle \nu_1 |\widehat{s}_z(\nu_1) \rangle\   &= (M - \lambda)^2 + |A|^2 - D^2=0 \label{spinWeylpoint3}\\
\langle \nu_2 |\widehat{s}_z(\nu_2) \rangle\   &= (M + \lambda)^2 + |A|^2 - D^2=0 \label{spinWeylpoint4}
\end{align}
and therefore $A=0$ and $M=0$. The SW points on each of the three non-trival phases
can be seen  in Table \ref{Table spin BHZ}. In this Hamiltonian the SW points
come in pairs with opposite chirality due to the TRS.

\begin{table*}[t]
\centering
\begin{tabular}{|| c | c|c |c|c |c||} 
 \hline
 Phase& Spin Weyl points & Spin Chern number & Spin Weyl indicator & FKM invariant & Spin invariants vector \\ 
 \hline\hline
\multirow{2}{6em}{$-3<\frac{M_0}{B_0} <1$} &  \multirow{2}{10em}{$(\pi,\pi, \pm\cos^{-1}( \frac{M_0}{B_0}-2))$} & $c_1^{s_z^-}(k_z=0)=0$ & \multirow{2}{4em}{SWI=1} & \multirow{2}{3em}{FKM=1} & \multirow{2}{5em}{$(1|000001)$} \\ 
& &  $c_1^{s_z^-}(k_z=\pi)=1$ & & & \\
 \hline
 \multirow{2}{6em}{$-1<\frac{M_0}{B_0} <1$} & $ (\pi,0, \pm\cos^{-1}( \frac{M_0}{B_0}))$ & $c_1^{s_z^-}(k_z=0)=-1$ &
 \multirow{2}{4em}{SWI=2} & \multirow{2}{3em}{FKM=0} & \multirow{2}{5em}{$(2|0000\minus 11)$} \\
 &  $(0,\pi, \pm\cos^{-1}( \frac{M_0}{B_0}))$ & $c_1^{s_z^-}(k_z=\pi)=1$ & & & \\ 
 \hline
\multirow{2}{6em}{$1<\frac{M_0}{B_0} <3$} &  \multirow{2}{10em}{$(0,0, \pm\cos^{-1}( \frac{M_0}{B_0}-2))$} & $c_1^{s_z^-}(k_z=1)=0$ &\multirow{2}{4em}{SWI=1}  & \multirow{2}{3em}{FKM=1} & \multirow{2}{5em}{$(1|000010)$} \\ 
& & $c_1^{s_z^-}(k_z=\pi)=0$  & &  &\\
 \hline
\end{tabular}
\caption{Topologically non-trivial phases of the 3D BHZ Hamiltonian. The SW points
appear on the four $k_z$ axis, and due to TRS, in pairs of opposite chirality. The first and the
third phase have non-trivial FKM invariant and SWI of 1, while the middle
phase has trivial FKM invariant with a SWI of 2. The vector
of spin invariants appears in the last column.}
\label{Table spin BHZ}
\end{table*}

Now let us find the linear expansion on ${\bf k}$ of the SV eigenvalues
around the SW points.
For this end we need to find an orthonormal basis of the valence states, and instead of doing it in complete generality, we will only calculate the linear $k_z$ expansion  centered on the point $(0,0, \cos^{-1}( \frac{M_0}{B_0}))$ for the phase $-1 < \frac{M_0}{B_0} < 1$.
\vspace{0.4cm}

Restricting the system to $k_x=0=k_y$ we find that the valence unitary eigenvectors for the Hamiltonian $H_{BHZ}({\bf k})$ for ${\bf k}=(0,0, \cos^{-1}( \frac{M_0}{B_0}))$  are:
\begin{align}
\psi_1& = \frac{(M-\lambda,-D,-(M-\lambda),D)}{\sqrt{2(M - \lambda)^2+2D^2)}}\\
\psi_2 &= \frac{(M-\lambda,D,M-\lambda,D)}{\sqrt{2(M - \lambda)^2+2D^2)}}.\\
\end{align}
The SV matrix becomes
\begin{align}
M_{{s}_z} = &\frac{(M - \lambda)^2-D^2}{(M - \lambda)^2+D^2} \ \sigma_x = -\frac{M}{\lambda } \ \sigma_x
\end{align}
where $\sigma_x$ is the Pauli matrix. Deriving with respect to $k_z$ and replacing $k_z=\cos^{-1}(\frac{M_0}{B_0})$
we obtain the $k_z$-linear term of the SV matrix:
\begin{align}
M_{{s}_z}  \sim \frac{B_0}{D_0}(k_z - \cos^{-1}(\tfrac{M_0}{B_0})) \ \sigma_x.
\end{align}
Note that when $D_0$ goes to zero,  the 
slope of the SV spectrum goes to infinity, and the anticommutator 
of the spin and the Hamiltonian goes to zero. This fact is further explored in Fig. \ref{shevssoc}.

Similar calculations can be performed for the $k_x$ and $k_y$ linear expansions, thus showing
that the SV eigenvalues are linear on $\mathbf{k}$ around the SW points

When the structural constants are $M_0=0$, $B_0=A_0=D_0=1$, one can show that 
\begin{align}
M_{{s}_z}  \sim k_x\sigma_z + k_y \sigma_y +(k_z - \cos^{-1}(\tfrac{M_0}{B_0}))  \sigma_x,
\end{align}
thus implying that the SW points have chirality $\pm1$ and that the SV operator
behaves like a $\bf{k} \cdot \bf{p}$ Hamiltonian. 

\subsection{Zeeman effect}

Consider the BHZ Hamiltonian subject to an external magnetic field in the spin direction
\begin{equation}
H({\bf k})  = H_{BHZ}({\bf k})  + {\mathrm{\mathbf B}}  \widehat{S}_z.
\end{equation}

The energy eigenvalues of the two valence states become
\begin{align}
\lambda &= - \sqrt{\pm 2\sqrt{ {\mathrm{\mathbf B}}^2(M^2+|A|^2)}+M^2+|A|^2  + D^2 +  {\mathrm{\mathbf B}}^2},
\end{align}
and one can see that there are Weyl type degenerate eigenstates whenever $|A|=0=M$. 
The same relations were found while solving equations (\ref{spinWeylpoint1}$-$\ref{spinWeylpoint4}) for the position of the SW points in the BHZ Hamiltonian. 
Therefore the SW points in the BHZ Hamiltonian, in the presence of a strong magnetic field aligned with the spin direction, 
evolve into energy Weyl points.

\subsection{Spin Weyl indicator}

The three nontrivial topological phases of the 3D BHZ Hamiltonian  
could be read from the amount of SW points present in the system.
The first and third phases provide examples of TIs with non-trivial FKM invariants,
thus making them strong TIs, while the second one has trivial FKM invariant
but its SW indicator is equal to 2.

The FKM invariant can be read from the eigenvalues of the Inversion operator
on the 8 TRIMs. The Inversion operator acts via the diagonal matrix
$\mbox{diag}(1,-1,1,-1)$ and the eigenvalues can be seen in Fig. \ref{phases BHZ}
The parity of the number of pairs of negative eigenvalues on the 8 TRIMs 
is the FKM invariant, and one can see in Fig. \ref{phases BHZ} that the second phase
has trivial FKM invariant. Calculating the first Chern class of the negative SV
eigenstates across the planes $k_z=0$ and $k_z=\pi$ we see that 
the absolute value of the difference of these Chern numbers is precisely the SWI. 
The information has been summarized in Table \ref{Table spin BHZ}.

\subsection{Spin Hall effect}

It is known that the spin Hall conductivity (SHC) within the energy band gap is a way to classify the charge-spin transport response in topological insulator materials. 
For 2D TI, the SHC takes a constant value within the band gap, and in ideal cases like the Kane-Mele model, it becomes quantized (QSHE) \cite{Kane-Mele-3dti}. 
In real materials, the inclusion of SOC induces spin mixing, breaking the commutativity between the spin and Hamiltonian operators, thereby resulting in a non-quantized value for the SHC within the energy gap \cite{SHE-in-2d-TI, SHE-in-3d-TI}. 
Despite this fact, the spin Chern numbers \cite{Prodan-SCN}, and therefore the SWI included in the last section, are well-defined quantities even in the presence of SOC. Therefore, the SWI is a 
quantity that permits to enhance the
characterization of  the underlying  properties of the system. 

\begin{figure}
	\includegraphics[width=8.2cm]{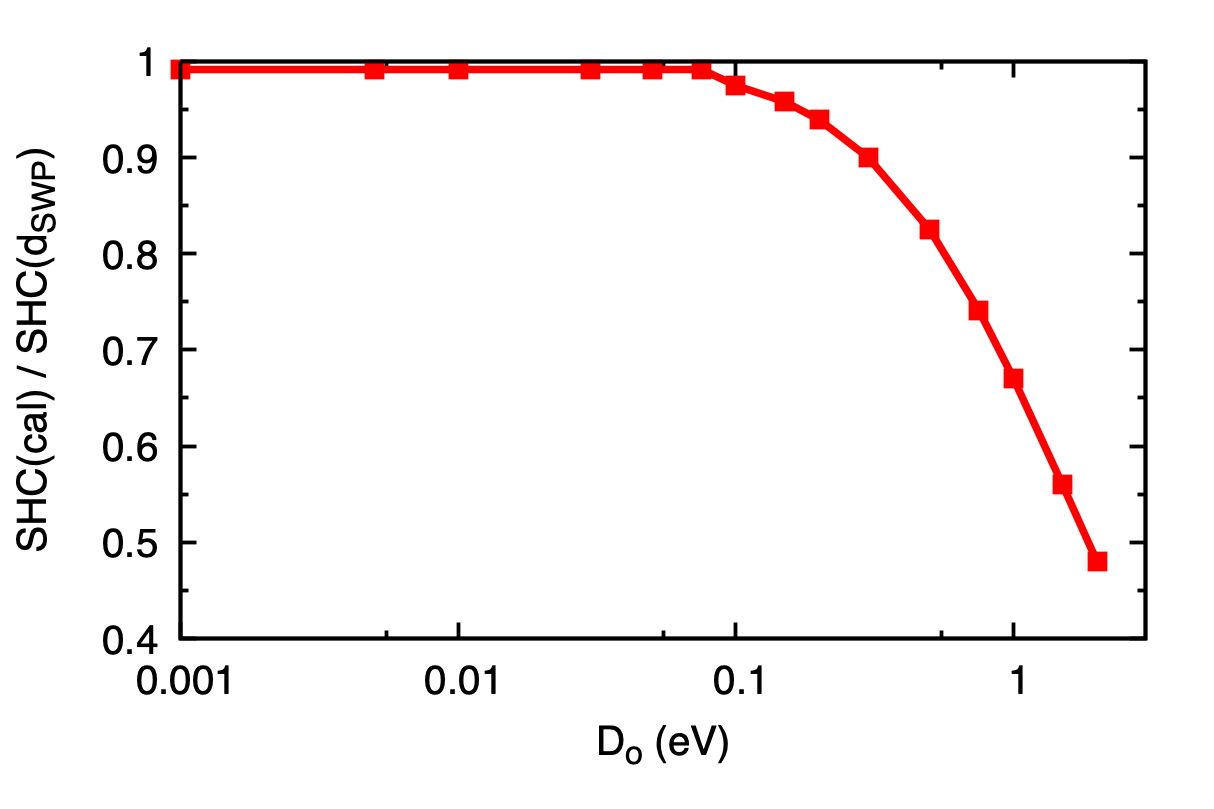}
	\caption{Spin Hall conductivity calculated with the Kubo formula  \eqref{Kubo} divided by the spin Hall conductivity calculated from the distance of the SW points in the reciprocal space \eqref{SHC SW} as a function of the SOC strength $D_0$ of the 3D BHZ model \eqref{BHZ Hamiltonian}. It is noted that the SHE inside the bang gap is proportional to the SW point distance in the negligible SOC limit.}
	\label{shevssoc}
\end{figure}

In 3D TIs, the characterization of non-trivial SHC within the band gap by topological
invariants is still an active area of research. 
The 3D BHZ Hamiltonian presented above offers a promising model to understand this relation. By performing SHC calculations for the 3D  BHZ model (see Fig. \ref{shevssoc}), we have found  that in the limit of minimal spin mixing (corresponding to a small $D_0$ term in the Hamiltonian \eqref{BHZ Hamiltonian}), 
the SHC becomes directly proportional to the proportion of $k$-layers  in the reciprocal space with spin Chern numbers equal to 1. 
This also permits to detect a relationship between the SHC and the distance between SW points, generalizing the well established relation between the  AHC and the distance between energy Weyl points \cite{ahe-in-weyls}. 
In the case of the two SW points in the third non-trivial phase of the 3D BHZ model, the SHC inside the band gap can be calculated as \cite{Felser-wp}:

\begin{equation} \label{SHC SW}
\sigma^z_{ij} =	-\frac{\hbar}{2e}\frac{e^2b_z}{2 \pi^2 \hbar}
\end{equation}
where $2b_z = 2 \cos^{-1}(\tfrac{M_0}{B_0}-2)$ is the distance between the SW points in the reciprocal space.  When considering the a finite spin mixing term in real materials, it becomes evident that the SHC presents a constant and non-quantized value within the energy gap, influenced by the strength of the SOC. This observation is depicted in Fig. \ref{shevssoc}.

\section{Tight binding model}

The spin Weyl indicator can also be incorporated in tight-binding (TB) models for 3D TIs. We have carried
out an extensive calculation on the TB model of the pristine pyrochlore model as was introduced by Varnava and Vanderbilt \cite{Vanderbilt-axion}. In this particular case  it is well known that the TB model
defines a 3D TI and its Hamiltonian has the following form:

\begin{align}
H = - t \mathop {\sum}\limits_{\langle {\boldsymbol{i}},{\boldsymbol{j}}\rangle, \sigma }  {\hat{\boldsymbol{c}}}_{\boldsymbol{i} \sigma}^\dagger {\hat{\boldsymbol{c}}}_{\boldsymbol{j} \sigma} 
 +  i \lambda \mathop {\sum} \limits_{\langle \langle {\boldsymbol{i,j}}\rangle \rangle,\alpha \beta} \nu_{ij}{\hat{\boldsymbol{c}}}_{\boldsymbol{i} \alpha}^\dagger {\boldsymbol{\sigma}_{\alpha \beta}} {\hat{\boldsymbol{c}}}_{\boldsymbol{j} \beta} 
\label{equa-tb}
\end{align}
where the first term represents the nearest-neighbor hopping interaction, while the second term represents
the  intrinsic SOC interaction (characterized by the coupling strength $ \lambda$). In this context, $\sigma _i$ represents the Pauli matrices, while ${\nu _{ij}}$ is determined by the cross product of $\boldsymbol{b_{ij}} \times \boldsymbol{d_{ij}}$ with
$\boldsymbol{d_{ij}}$. Here $\boldsymbol{d_{ij}}$ is the unit vector connecting site $i$ with site $j$, and 
$\boldsymbol{b_{ij}}$ is the unit vector
from the center of a tetrahedron to the midpoint of the bond
$\langle ij \rangle$. This model is exactly the one that appears in the reference \cite[\S 3]{Vanderbilt-axion}. .

\begin{figure}
	\includegraphics[width=8.6cm]{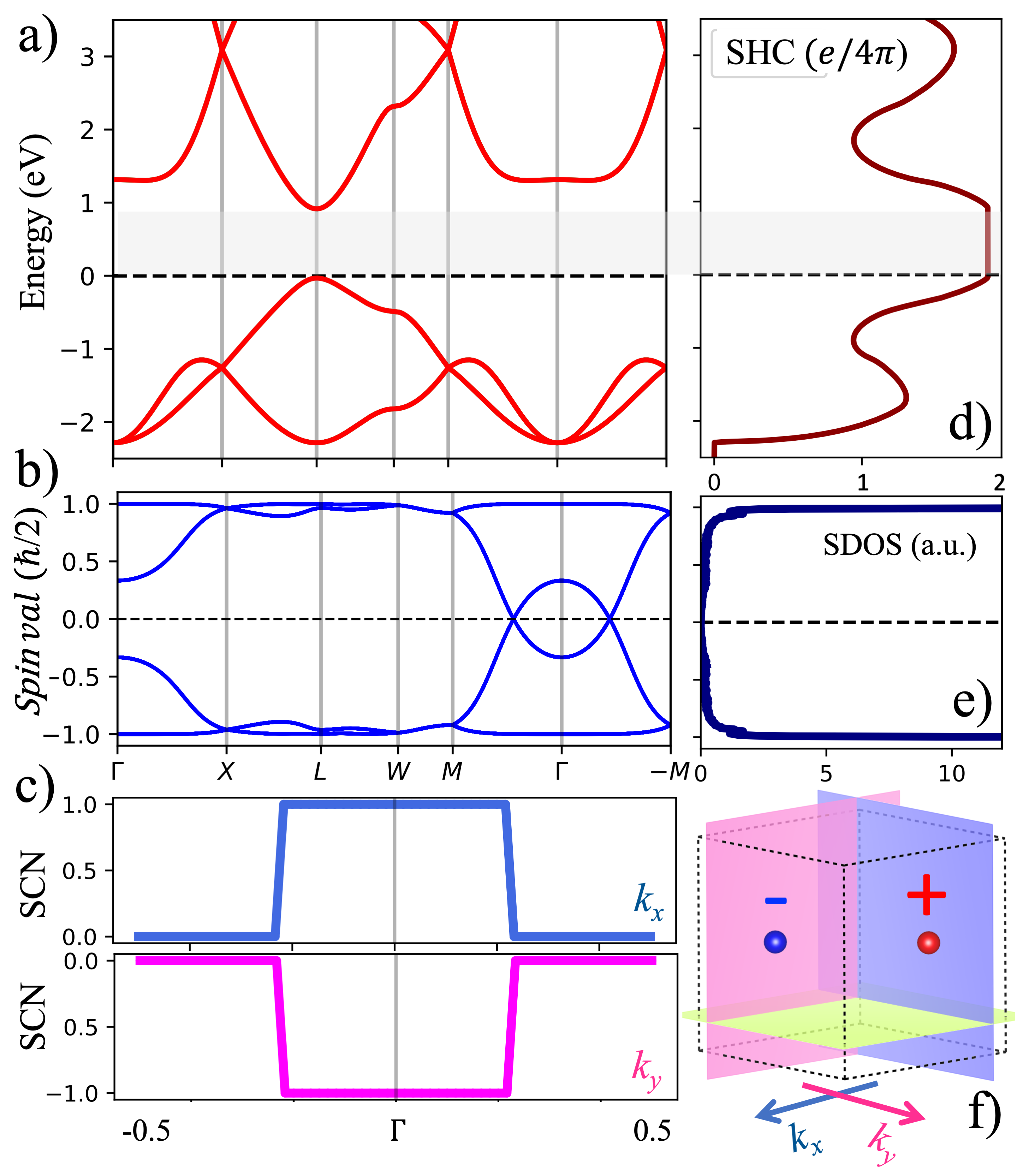}
	\caption{Tight-binding model introduced in \cite{Vanderbilt-axion} with the Hamiltonian of Eqn. \eqref{equa-tb} and 
$\lambda$=0.3$t$.
(a) Bulk band structure and (b) spin valence spectrum along high symmetry lines in the BZ. (c) spin Chern number (SCN) calculated for perpendicular planes along the $k_x$ and $k_y$ reciprocal axis, (d) spin Hall conductivity and (d) spin density of states as a function of the Fermi energy of the tight-binding model. f) Position of the spin Weyl points in the reciprocal space with the $k$-planes used for the SCN calculation.}
	\label{modelo-tb}
\end{figure}

This model exhibits a 3D TI that maintains time-reversal and inversion symmetry. By setting the parameter $ \lambda$=0.3$t$, we computed the band structure at half-filling, revealing a bandgap of approximately 1.0 eV, as depicted in Fig. \ref{modelo-tb}. We observed the emergence of  two SW points along the -M-$\Gamma$-M path in the BZ, indicating the presence of these novel spin topological indicator.  TB calculations along different $k$-paths revealed that the energy and SV spectrum are gapped in other regions of the BZ. 

This system exhibits a spin Chern number transition along planes perpendicular
 to both the $k_x$ and the $k_y$ axis. The presence of SW points
 in the bulk is inferred from the SCN signal along the $k_x$ and $k_y$ axis
 as shown in Fig. \ref{modelo-tb} c). Fig. \ref{modelo-tb} f) depicts the position of the SW points with opposite chirality
 that produces the SCN signal in the system.
This system models a SWTI with SWI=1, confirming the Strong Topological
Insulating property of the pyrochlore lattice shown in \cite{Guo_pyrochlore, Moyuru_pyrochlore}.
 The spin Invariants vector for the TB model is $(1|10\minus 1000)$.

\section{Materials realization}

\begin{figure*}
	\includegraphics[width=16.2cm]{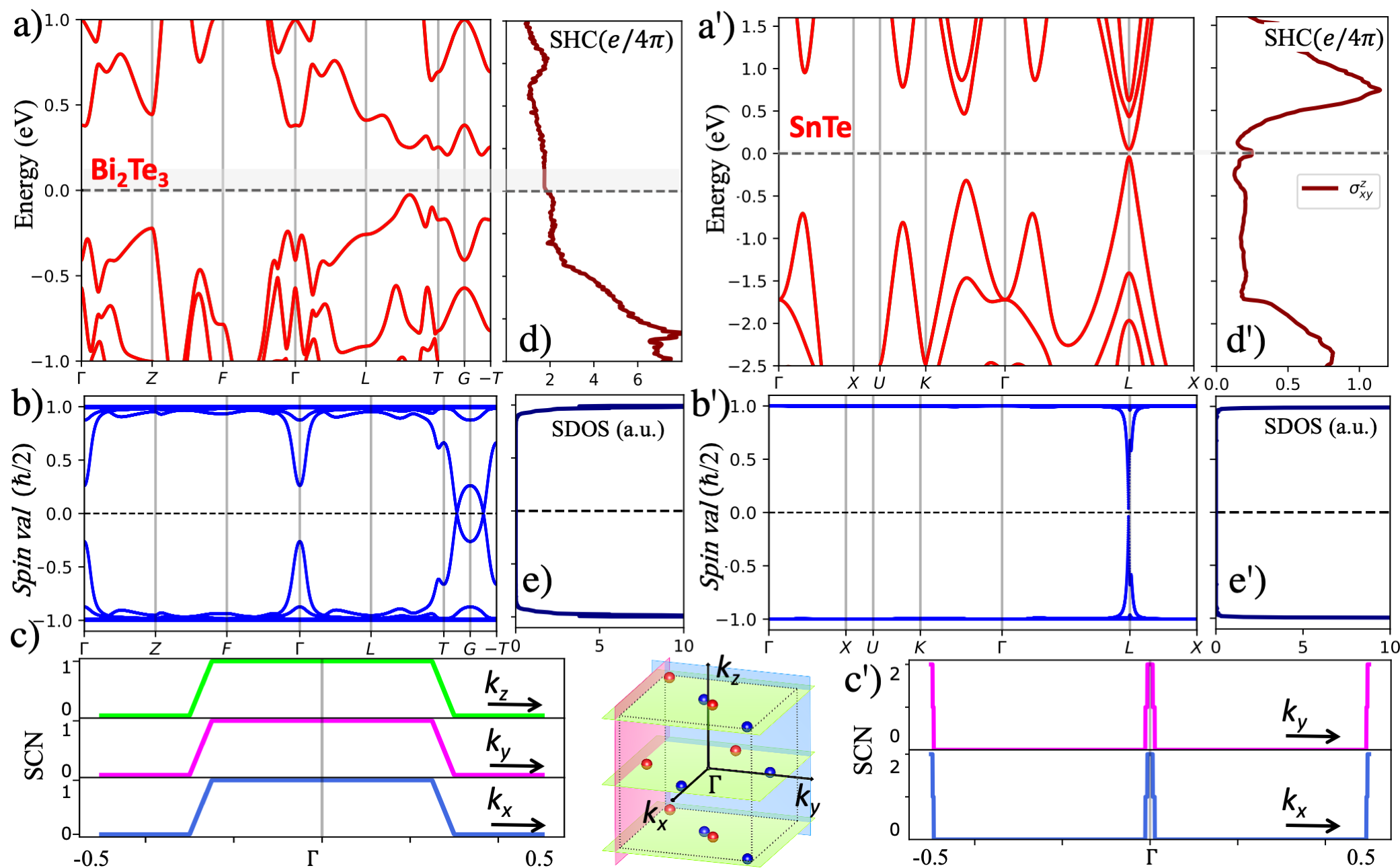}
	\caption{(a) and (a') Bulk band structure, (b) and (b') spin valence spectrum along high symmetry lines in the BZ, (c) and (c') spin Chern number -SCN- calculated for perpendicular planes along the $k_x$, $k_y$ and $k_z$ reciprocal axis, (d) and (d') spin Hall conductivity, (e) and (e') spin density of states as a function of the Fermi energy of the 3D topological materials Bi$_2$Te$_3$ and SnTe. The central panel at the bottom
presents the position of the eight SW points of SnTe with their chiralities.}
	\label{modelo-bi2se3}
\end{figure*}

We have calculated the spin invariants presented above in real materials, and we have
focused our attention to Bi$_2$Te$_3$ which is a 3D STI, and and SnTe,
which exhibits a distinct spin Weyl topological insulating property.
The first material is modeled in a rhombohedral unit cell and consist of two layers of Bi atoms and three layers of Te/Se atoms, arranged in a quintuple layer structure. The coupling between atomic layers within one quintuple layer is strong, but much weaker between two quintuple layers  \cite{Chen-bi2te3}. The electronic band structure of Bi$_2$Te$_3$ in the rhomboidal crystal structure is shown in Fig. \ref{modelo-bi2se3}a).  The material exhibits an indirect band gap energy of approximately 0.2 eV.

In the SV spectrum shown in Fig. \ref{modelo-bi2se3}b) we observe the presence of two SW points
along the T-$\Gamma$-T $k$-path, corresponding to the main diagonal of the BZ. 
As depicted in Fig. \ref{modelo-bi2se3}e), the spin density of states reveals that the eigenvalues of the spin valence operator are concentrated around $\pm$1 values, except at the SW points.
These SW points exhibit opposite chirality and give rise to a transition of the SCN when scanned across perpendicular $k$-planes in the reciprocal lattice.
From Fig. \ref{modelo-bi2se3}c) it is noted that the SCN changes precisely at the positions of the SW points, signifying a topological transition between distinct planes of reciprocal space.

The topological nature of Bi$_2$Te$_3$ is evident in our SHC calculation, where a non-zero signal in Fig. \ref{modelo-bi2se3}d) is observed within the band gap. 
This material can be classified by a SWI=1, in agreement with the Fu-Kane-Mele invariant of 1. This indicator is obtained from the calculation of Wilson loops on the $k_l$=0, $\pi$ planes, corresponding to 0 and 1 respectively. 
 The position and chirality of the SW points can be inferred from the change of the topological index of 0 at $k_l$=0 to 1 at $k_l$=$\pi$ along each $k_l$ direction (See Fig. \ref{modelo-bi2se3}c), and 
 its spin invariant vector becomes $(1|101010)$.

On the other hand, the SnTe exhibits a rocksalt crystal structure with two atoms by unit-cell \cite{Hsieh-snte}, and its band gap of around 0.1 eV is located at four equivalent L points within the face-centered-cubic BZ as can be seen in Fig. \ref{modelo-bi2se3}a'). Despite the trivial FKM  invariant calculation predicting a trivial character for SnTe, this material was previously classified as a crystalline topological insulator based on the presence of a non-trivial mirror Chern number \cite{Hsieh-snte}. Our findings have been corroborated by the observation of a non-zero SHC within the band gap, as shown in Fig. \ref{modelo-bi2se3}d').

From the SV spectrum presented in Fig. \ref{modelo-bi2se3}b'), eight SW points are detected close to the L and T points inducing a change of SCN as presented in Fig. \ref{modelo-bi2se3}c'). Accordingly, this material is classified by a SWI=4.
The existence of these SW points serves as an indication of the transition in the internal topological phase along the $k_x$, $k_y$ and $k_z$ directions. 
This transition involves a shift from a spin Chern number of 2 at the planes $k_x=0$ and $k_y=0$ to 0 in the interior of the interval $(0, \pi)$, to again a shift back to 2 across
the planes $k_x=\pi$ and $k_y=\pi$. This feature confirms the SWTI nature of SnTe, but more importantly, it highlights a significant difference from the conventional classification of SnTe using the Fu-Kane-Mele invariant. Here is worth mentioning that
knowing the spin Chern number of the system along the planes $k_l$=0,$\pi$ is not enough
to distinguish its topological nature. In the particular case of SnTe, the spin Chern numbers
along the planes $k_l$=0,$\pi$ for $l=x,y$ is 2, while along the planes $k_z$=0,$\pi$ is 0.
One could mistake this material as a SCI, since the spin Chern numbers are equal along
parallel planes $k_l$=0,$\pi$. Nevertheless, its correct classification is being an SWTI with
spin invariants vector equal to $(4|222200)$.

Finally, Table \ref{Table_Materials} presents a summary of the 3D trivial and topological insulator materials studied in this work.
The table includes the number of spin Weyl points, the spin Weyl indicator, the Fu-Kane-Mele invariant, the spin invariants vector, and the respective (time-reversal and inversion) symmetries associated with each material.
We have considered representative examples from different 3D insulator materials to highlight
the classification presented in Fig. \ref{classes}. 
Te, GaAs, AuF$_3$, and CaMnO$_3$ display trivial insulator behavior with energy and spin gaps. 
In contrast, Bismuth (Bi) exhibits a constant spin Chern number of 2 along three directions in the BZ, indicating its classification as a spin Chern insulator.  This is consistent with previous theoretical investigations that have demonstrated the presence of a 3D topological band structure in Bi \cite{Schindler-bihoti}.

The spin Weyl indicator confirms that the material Bi$_2$Se$_3$ exhibits a strong topological character similar to Bi$_2$Te$_3$, consistent with previous theoretical and experimental studies \cite{Zhang-bi2se3,Chen-bi2te3}.
Finally, material SnTe exhibits a spin Weyl topological insulator property
having eight SW points in the bulk; and whose distribution can be seen in the central-bottom panel of Fig. \ref{modelo-bi2se3}. This material has a trivial FKM invariant but Fig. \ref{modelo-bi2se3}d') shows that it  exhibits a non-zero SHC inside the band gap. Therefore we claim that the SWI is more suited to detect SHC signals in 3D topological insulator materials.

Here it is worth mentioning that the properties of the spin Chern number and spin Weyl points in topological insulators remain valid when TRS invariance is broken. This is shown in the case of CaMnO$_3$ and MnBi$_2$Te$_4$, where the collinear antiferromagnetic phase with Neel vector along the $z$-axis was considered (see Table \ref{classes}). The topological invariant vector confirms the predicted topological response in MnBi$_2$Te$_4$, as reported in previous works \cite{Deng-mnbi2te4,li-mnbi2te4}.

\begin{table*}[t]
	\centering
	\begin{tabular}{||c|c|c|c|c|c|c|c||} 
		\hline
		Material & SW points & SWI  & FKM Inv. & Spin inv. vec.&TRS &Inversion symm.  & Spin top. class. \\ 
		\hline\hline
		Te           & 0 & 0 & 0 & $(0|000000)$  & \checkmark & $X$& SI\\ 
		GaAs      & 0 & 0 & 0 &$(0|000000)$ &\checkmark & $X$ & SI\\
		AuF$_3$ & 0 & 0 & 0 &$(0|000000)$ &\checkmark & $X$ & SI\\
		Bi         \cite{Schindler-bihoti}         & 0 & 0 & 0 &$(0|222222)$ &\checkmark & \checkmark & SCI\\ 
		Bi$_2$Te$_3$   \cite{Chen-bi2te3}   & 2 & 1 & 1 & $(1|101010)$ &\checkmark & \checkmark & STI\\ 
		Bi$_2$Se$_3$    \cite{Zhang-bi2se3}  & 2 & 1 & 1 & $(1|101010)$ &\checkmark & \checkmark & STI\\ 
		SnTe                   \cite{Hsieh-snte}    & 8 & 4 & 0 & $(4|222200)$ &\checkmark & \checkmark & SWTI\\ 
		CaMnO$_3$ & 0 & 0 & 0 &$(0|000 000)$ & $X$   &\checkmark& SI\\ 
		MnBi$_2$Te$_4$ \cite{Deng-mnbi2te4,li-mnbi2te4} & 2 & 1 & 1 &   $(1|\minus1010\minus10)$& $X$  &\checkmark & STI\\ 
		\hline
	\end{tabular}
	\caption{spin topology calculated for 3D materials. spin Weyl points, spin Weyl indicator, Fu-Kane-Mele index, spin invariants vector, time-reversal and inversion symmetry, and spin topology classification according to Fig. \ref{classes}. Bismuth (Bi) has a constant spin number of 2 along three directions of the BZ. MnBi$_2$Te$_4$ and CaMnO$_3$ were calculated in the collinear antiferromagnetic phase with Neel vector along the $z$-axis.}
	\label{Table_Materials}
\end{table*}

\section{Computational details}

\begin{figure}
	\includegraphics[width=8.6cm]{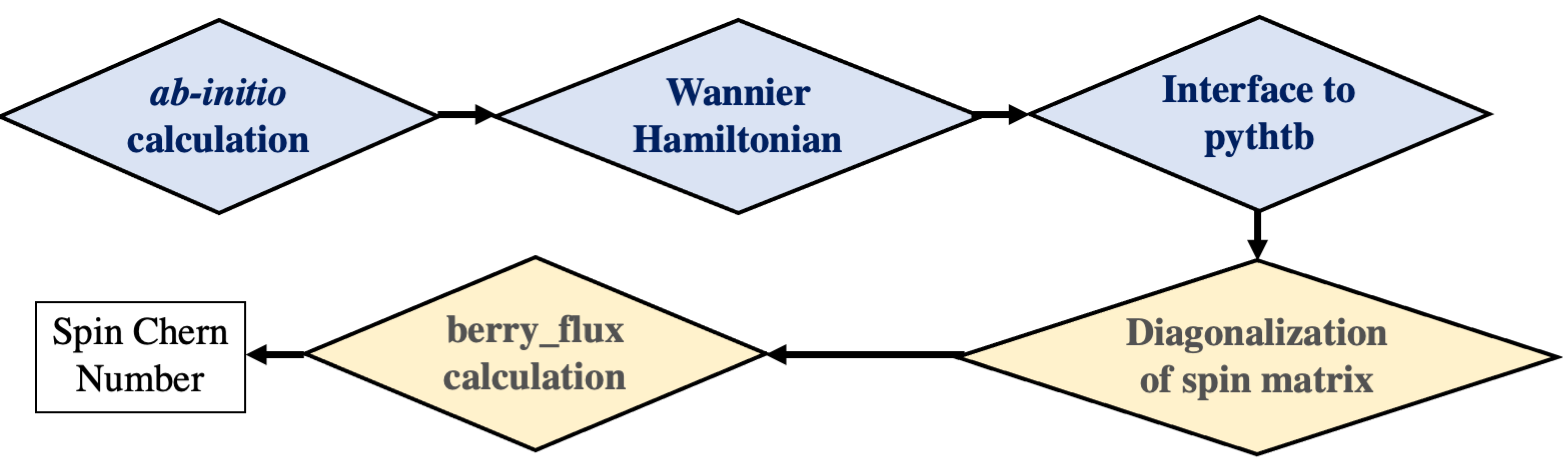}
	\caption{Workflow showing the computational method. Ab-initio calculations (VASP code) are used to find the electronic structure of the material. Subsequently, the \texttt{wannier90} code is utilized to construct the wannier Hamiltonian, which serves as a basis for the generation of the tight-binding model using the \texttt{pythtb-wannier90} interface. The next step involves the generation of the spin matrix operator for valence states, followed by its diagonalization to obtain the spin valence spectrum. Finally, the Berry curvature is integrated over 2D $k$-planes to determine the spin Chern number.}
	\label{method}
\end{figure}  

We used density-functional theory (DFT) calculations to investigate the spin topology of magnetic and nonmagnetic materials. The generalized gradient approximation (GGA) \cite{pbe}, as implemented in the Vienna \textit{ab-initio} simulation package (\texttt{VASP}) \cite{vasp}, was used to account for exchange and correlation effects. 
In the electronic structure calculations, we expanded the electron wave function in plane waves with a cutoff energy of 520 eV. The Brillouin zone was sampled using a $k$-mesh of 0.03 (2$\pi$/\AA) $k$-space resolution. 
The lattice constants for the studied materials were obtained from the Materials Project database \cite{materialsproject}.

We employed the \texttt{wannier90} code to build the maximally localized Wannier basis \cite{wannier90} as a post-processing approach following the DFT calculations. 
The \texttt{pythtb-wannier90} interface of the \texttt{pythtb} code \cite{pythtb} was utilized to generate tight-binding Hamiltonians for the TB model and for each material. 
Next, we used the electron wave functions of the valence states to generate the spin valence matrix operator and performed its diagonalization to obtain the spin valence spectrum. 
To study the spin topological properties, we used the spin valence eigenvectors to integrate the Berry curvature (\texttt{berry-flux} utility in \texttt{pythtb}) \cite{pythtb} over 2D $k$-planes, thereby extracting spin Chern numbers across the Brillouin Zone. 
The workflow of this method is shown in Fig. \ref{method}.

The intrinsic spin Hall conductivity was calculated using the \texttt{WannierBerri} code \cite{wannierberri} by integrating the spin Berry curvature over the first Brillouin zone. For the case of SHC ($\sigma^z_{xy}$) we set:

\begin{equation} \label{Kubo}
\sigma^z_{xy}=-\dfrac{e^2}{\hbar} \sum_{n}\int_{BZ}\dfrac{dk^3}{(2\pi)^3}f_n(k)\Omega^z_{xy}(k)
\end{equation}
where $f_n(k)$ is the Fermi-Dirac distribution and the Berry curvature $\Omega^z_n(k)$ for the $n$th band is:

\begin{equation}
\Omega^z_{xy}(k)= -2{\hbar}^2 \text{Im}\sum_{m\neq n}%
\frac{\left\langle  \psi_n \vert \widehat j^z_{x} \vert \psi_m \rangle\langle \psi_m \vert \widehat v_y \vert \psi_n \right\rangle}{\left(\epsilon_{n,k}-\epsilon_{m,k}\right)^{2}} 	
\end{equation}
where $\vert \psi_n(k) \rangle$ are the Bloch functions of a single band $n$, $k$ is the Bloch wave vector, $\epsilon_{n,k}$ is the band energy,  $\hat{v}_i$ is the velocity operator in the $i$ direction and 
$\hat{j}^z_x=\frac{1}{2}\{\hat{v}_x,\hat{s}_z\}$ is the spin current operator.  It is important to note that the $\hat{j}^z_x$ definition does not incorporate the spin torque contributions, which is taken into account when the universal spin current operator is used \cite{PhysRevLett.96.076604}.
Finally, the FKM index was computed using the \texttt{WannierTools} code \cite{wanniertools}.

\section{Conclusions}

We have found that 3D topological insulators (TI) can be identified by the presence of spin Weyl points in their spin valence spectrum or by the non-triviality of the spin Chern numbers. Both indicators serve as novel predictors  of topological insulating phases.  
It is important to note that these phenomena cannot be regarded simply as a stack of two-dimensional states, thus making these indicators truly 3-dimensional.
In addition, we have found a correlation between the presence of spin Weyl points in 3D TIs and the topological signal of the spin Hall effect (SHE) for the 3D BHZ model.

We propose the use of the spin invariants vector to enhance the topological classification of 3D materials.
This vector contains the information regarding the spin Weyl indicator, and the values of the spin Chern numbers of the negative spin valence eigenvectors across the planes $k_l$=0,$\pi$ for $l$=$x,y,z$.
This vector contains the necessary information to distinguish the material as a spin insulator, 
a spin Chern insulator or a spin Weyl topological insulator. We have provided an array of materials
which exhibit interesting spin topology phases and whose spin invariants vector predicts its topological nature.

The novelty of the spin invariants vector is shown when applied to Bismuth and SnTe. In the former
case the material exhibits a spin Chern insulating property, while the former shows
a spin Weyl topological insulating property. In both cases, the new invariant permits to carry out a 
precise classification of these with regards to their spin properties. 

We believe that the spin invariants vector adds information to the characterization of the topological classification of materials,
enhancing the known invariants and detecting new spin topological phases. Nevertheless, the relation of the spin invariants vector with the whole package of electromagnetic properties stills needs to be investigated.

\section*{Acknowledgments}
The first author gratefully acknowledges the computing time granted on the supercomputer Mogon at Johannes Gutenberg University Mainz (hpc.uni-mainz.de). 
The second author acknowledges the support of CONACYT through project CB-2017-2018-A1-S-30345-F-3125 and of the Max Planck Institute for Mathematics in Bonn, Germany. 
Both authors thank the continuous support of the Alexander Von Humboldt Foundation, Germany.

\bibliographystyle{unsrt}
\bibliography{topological}

\end{document}